\documentclass[twocolumn,nofootinbib,showpacs,prl]{revtex4}
\usepackage{graphicx}
\usepackage{color}
\begin{document}
\title{Evidence for further $c\bar{c}$ vector resonances}
\author{Eef van Beveren}
\email{eef@teor.fis.uc.pt}
\affiliation{Centro de F\'{\i}sica Computacional,
Departamento de F\'{\i}sica, Universidade de Coimbra,
P-3004-516 Coimbra, Portugal}
\author{George Rupp}
\email{george@ist.utl.pt}
\affiliation{Centro de F\'{\i}sica das Interac\c{c}\~{o}es Fundamentais,
Instituto Superior T\'{e}cnico, Universidade T\'{e}cnica de Lisboa,
Edif\'{\i}cio Ci\^{e}ncia, P-1049-001 Lisboa, Portugal}
\date{\today}

\begin{abstract}
We discuss the shape of threshold signals in
production cross sections of the reaction
$e^{+}e^{-}\to D^{\ast}\bar{D}^{\ast}$,
at the opening
of the $D^{\ast}_{s}\bar{D}^{\ast}_{s}$
and $\Lambda_{c}^{+}\Lambda_{c}^{-}$ channels.
Furthermore, evidence for the
$\psi (3D)$, $\psi (5S)$, $\psi (4D)$, $\psi (6S)$,
$\psi (5D)$, $\psi (7S)$, $\psi (6D)$, and $\psi (8S)$
new charmonium vector resonances is presented, on the basis of
data recently published by the BaBar Collaboration.
Central masses and resonance widths are estimated.
Confirmation of these resonances would be a huge step in lifting the
precision level of hadron spectroscopy towards that of atomic spectroscopy,
with far-reaching consequences for theory.
\end{abstract}


\maketitle

\begin{figure}[htbp]
\begin{center}
\begin{tabular}{c}
\includegraphics[height=150pt]{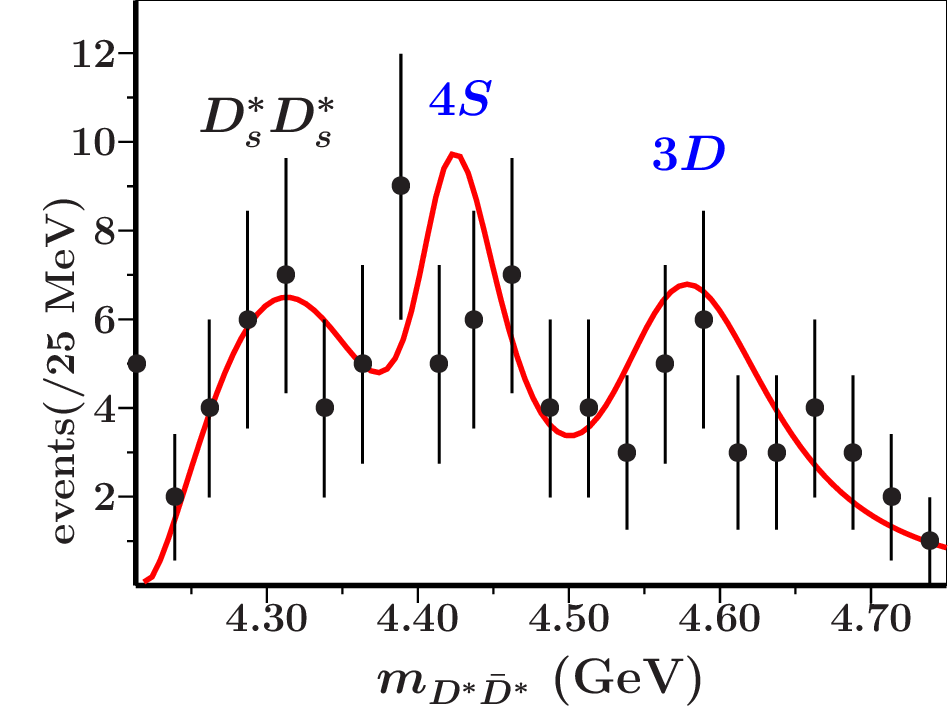}\\ [-10pt]
\end{tabular}
\end{center}
\caption{\small
Event distribution,
as published by the BaBar Collaboration
in Ref.~\cite{PRL101p172001},
for the reaction $e^{+}e^{-}\to D^{\ast}\bar{D}^{\ast}$.
}
\label{B4260b}
\end{figure}

Recent data published by the BaBar Collaboration~\cite{PRD79p092001}
do not exhibit the $X(4260)$ \cite{PLB667p1} structure in
$e^{+}e^{-}\to D^{\ast}\bar{D}^{\ast}$
(see Fig.~\ref{B4260b}).
However, the data clearly show an enhancement due to the opening
of the $D^{\ast}_{s}\bar{D}^{\ast}_{s}$ channel at 4.213 GeV.
\begin{figure}[htbp]
\begin{center}
\begin{tabular}{c}
\includegraphics[height=150pt]{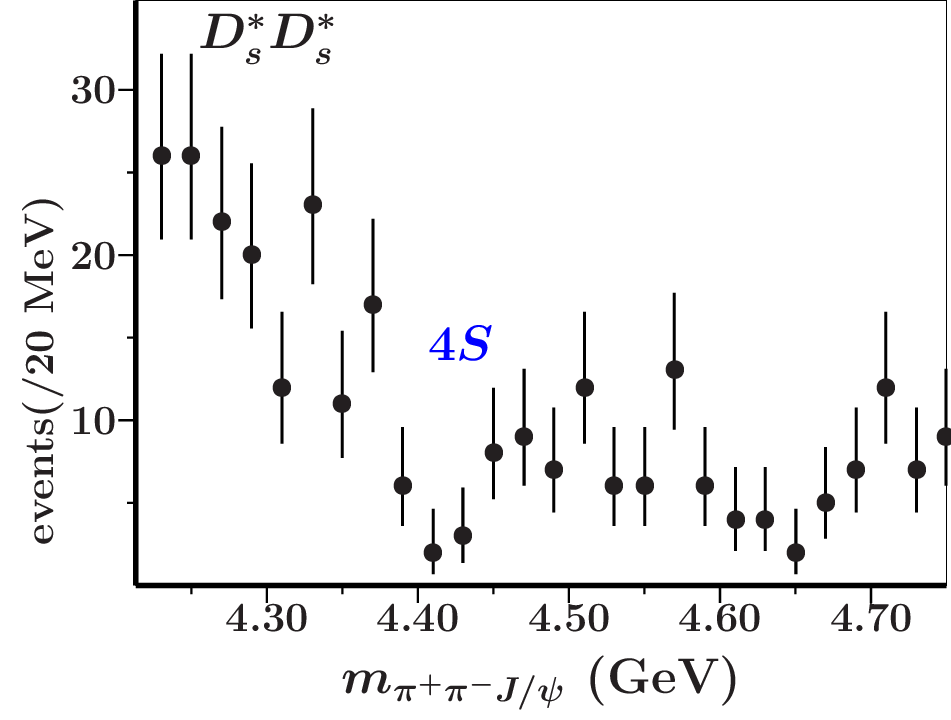}\\ [-10pt]
\end{tabular}
\end{center}
\caption{\small
Event distribution,
as published by the BaBar Collaboration
in Ref.~\cite{PRL95p142001},
for the reaction $e^{+}e^{-}\to\pi^{+}\pi^{-}J/\psi$.
}
\label{B4260a}
\end{figure}
The $X(4260)$ $J^{PC}=1^{--}$ charmonium enhancement,
discovered in $\pi^+\pi^-J/\psi$ by the BaBar Collaboration
\cite{PRL95p142001} (see Fig.~\ref{B4260a}) and
originally baptized as $Y(4260)$, was later confirmed and also seen in
$\pi^0\pi^0J/\psi$ as well as $K^+K^-J/\psi$ by the CLEO Collaboration
\cite{PRL96p162003}, and finally by the Belle Collaboration,
in $\pi^+\pi^-J/\psi$ \cite{PRL99p182004}, too. Moreover, both BaBar and
Belle observed a structure in $e^{+}e^{-}\to\pi^{+}\pi^{-}\psi(2S)$ at
somewhat higher energies, namely at 4.32~GeV \cite{PRL98p212001} and
4.36~GeV \cite{PRL99p142002}, respectively. According to BaBar
\cite{PRL98p212001}, their very broad enhancement at 4.32~GeV might just
correspond to a different decay mode of the $X(4260)$. On the other hand,
the much narrower Belle structure at 4.36~GeV, while not incompatible
with the latter BaBar state, seems more difficult to reconcile with the
$X(4260)$.

In Fig.~\ref{B4260b} we indicate by a solid line
our interpretation of the data of Ref.~\cite{PRD79p092001}
just above the $D^{\ast}_{s}\bar{D}^{\ast}_{s}$ threshold.
One clearly observes --- albeit with very limited statistics ---
a threshold enhancement, as predicted by us in Ref.~\cite{AP323p1215},
as well as the two $c\bar{c}$ resonances $\psi(4S)$ and $\psi(3D)$.
The latter charmonium state can be determined from the theoretical model
of Ref.~\cite{PRD21p772}, and was also predicted by Godfrey and Isgur
\cite{PRD32p189}, though a little bit lower, viz.\ at 4.52~GeV.
The $D^{\ast}_{s}\bar{D}^{\ast}_{s}$ threshold enhancement rises fast
and peaks at about 4.32 GeV. For higher invariant masses,
the threshold signal slowly decreases, almost vanishing at about 4.75 GeV.

In Ref.~\cite{IJTPGTNO11p179}, we derived a precise relation between the
formalism of non-exotic meson-meson scattering due to a resonating
$s$-channel quark-antiquark propagator in the intermediate state, and the
deformed $q\bar{q}$ resonance spectrum owing to the inclusion of infinite
chains of meson loops.
Moreover, in Ref.~\cite{AP323p1215} we deduced an amplitude for production
processes, resulting in a complex relation \cite{EPL81p61002}
between production and scattering amplitudes.
The latter relation is formally equivalent \cite{EPL84p51001}
to the real relation of Au, Morgan, and Pennington \cite{PRD35p1633},
but with an important difference:
whereas the coefficients of the complex relation \cite{EPL81p61002}
are of a purely kinematical origin,
the real coefficients of Ref.~\cite{PRD35p1633}
contain the scattering amplitudes themselves
\cite{EPL84p51002}.
As a consequence, one does not find
a distinct threshold enhancement
in the formalism of Ref.~\cite{PRD35p1633}.

The question of interest here is:
why is the signal in $e^{+}e^{-}\to\pi^{+}\pi^{-}J/\psi$
depleted exactly at the mass of the $\psi (4S)$?
In Refs.~\cite{HEPPH0605317,ARXIV08111755,ARXIV09044351},
we have discussed this issue and come to the following conclusion:
while the reaction $e^{+}e^{-}\to\pi^{+}\pi^{-}J/\psi$ is dominated
\begin{figure}[htbp]
\begin{center}
\begin{tabular}{cc}
\includegraphics[height=150pt]{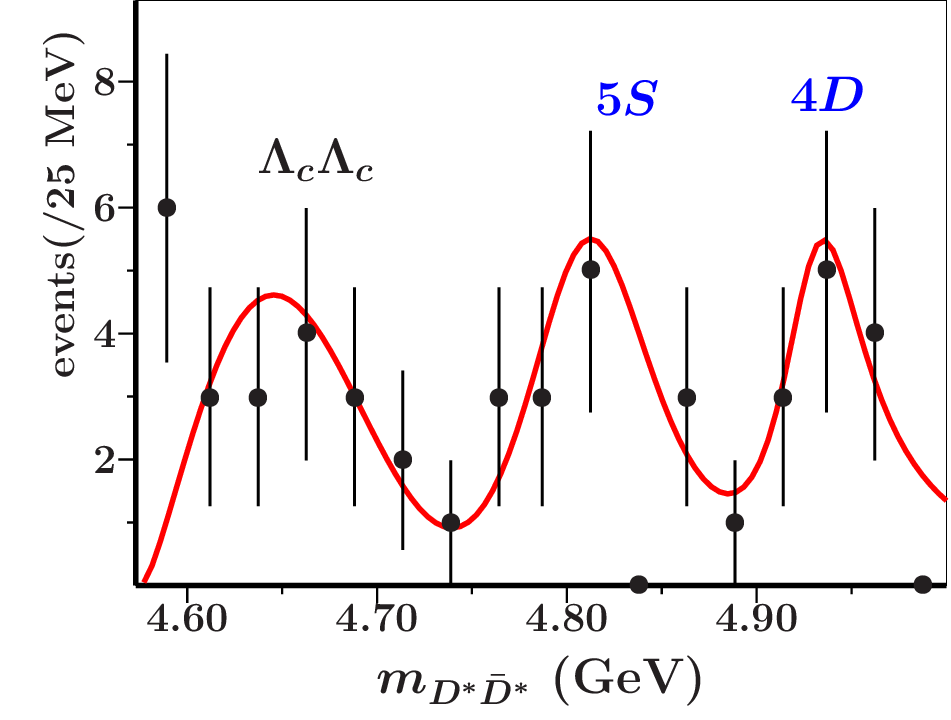}\\ [-10pt]
\end{tabular}
\end{center}
\caption{\small
Event distribution for the reaction
$e^{+}e^{-}\to D^{\ast}\bar{D}^{\ast}$,
obtained by the BaBar Collaboration \cite{PRD79p092001},
near the $\Lambda_{c}^{+}\Lambda_{c}^{-}$ threshold.
One observes signals of the $\psi (5S)$
and $\psi (4D)$ resonances.
The threshold enhancement for $\Lambda_{c}^{+}\Lambda_{c}^{-}$
is less pronounced than in the reaction
$e^{+}e^{-}\to\Lambda_{c}^{+}\Lambda_{c}^{-}$
(see Fig.~\ref{LcLca}).
}
\label{LcLcb}
\end{figure}
by a peripheral, OZI-forbidden process,
in which a $\sigma$-like structure, i.e., $f_{0}(600)$ and/or
$f_{0}(980)$, is radiated off by the gluon cloud,
the reaction $e^{+}e^{-}\to D^{\ast}\bar{D}^{\ast}$
is dominated by OZI-allowed quark-pair creation in the inner core
of the $c\bar{c}$ propagator.
Near a resonance of the $c\bar{c}$ propagator,
the latter --- faster --- process dominates,
hence depleting the $\pi^{+}\pi^{-}J/\psi$ signal
(see Fig.~\ref{B4260a}).
The $X(4260)$ enhancement is probably caused by the fact that
in an $s\bar{s}$-rich environment,
which stems from $D^{\ast}_{s}\bar{D}^{\ast}_{s}$ formation with
sufficient phase space, a relatively stable $f_{0}(980)$
can be formed.
These two processes of different origins, and with different frequencies,
may also give rise to interference patterns \cite{PRD79p111501R}.

\begin{figure}[htbp]
\begin{center}
\begin{tabular}{c}
\includegraphics[height=150pt]{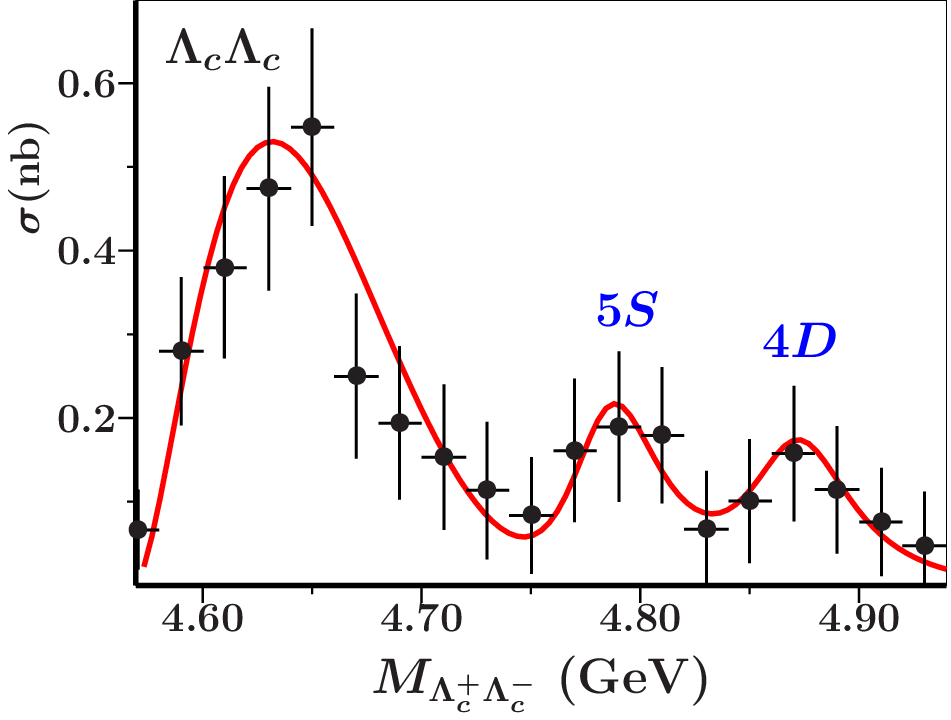}\\ [-10pt]
\end{tabular}
\end{center}
\caption{\small
Experimental cross section
for the reaction
$e^{+}e^{-}\to\Lambda_{c}^{+}\Lambda_{c}^{-}$,
obtained by the Belle Collaboration \cite{PRL101p172001},
near the $\Lambda_{c}^{+}\Lambda_{c}^{-}$ threshold.
One observes signals of the $\psi (5S)$
and $\psi (4D)$ resonances.
The threshold enhancement for $\Lambda_{c}^{+}\Lambda_{c}^{-}$
is more pronounced than in the reaction
$e^{+}e^{-}\to D^{\ast}\bar{D}^{\ast}$
(see Fig.~\ref{LcLcb}).
}
\label{LcLca}
\end{figure}

\begin{figure}[htbp]
\begin{center}
\begin{tabular}{c}
\includegraphics[width=244pt]{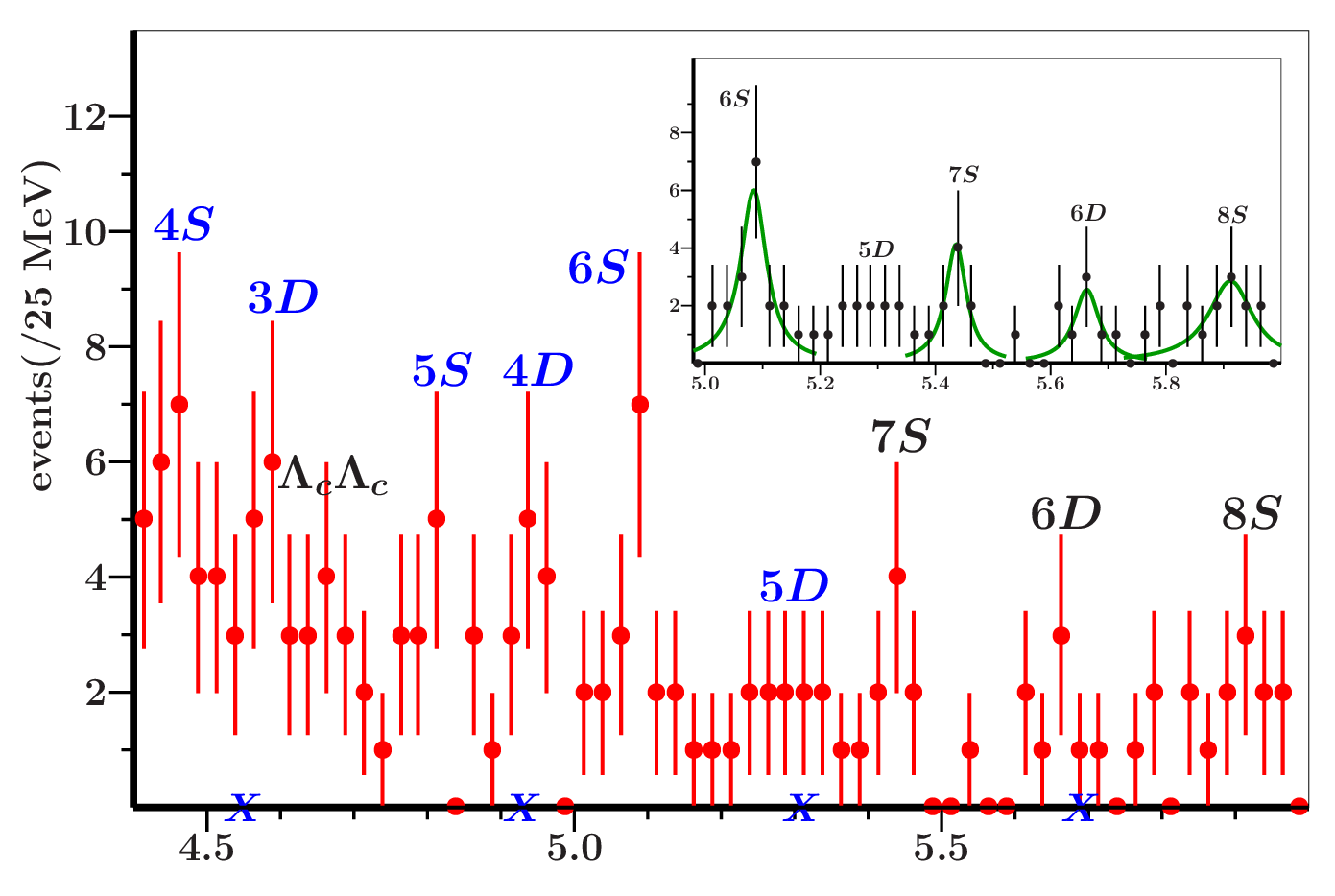}
\end{tabular}
\end{center}
\caption{\small
Event distribution in the invariant-mass interval 4.4--6.0~GeV,
for the exclusive production of $D^{\ast}\bar{D}^{\ast}$
in initial-state-radiation events, from $e^{+}e^{-}$ annihilations
at a center-of-mass energy near 10.58~GeV,
as published by the BaBar Collaboration \cite{PRD79p092001}.
With ({\color{blue} X}) on the horizontal axis we indicate
the harmonic-oscillator vector levels for the parameters
$m_{c}=1.562$ GeV and $\omega =0.19$ GeV \cite{PRD27p1527}.
Meson loops, first introduced in Ref.~\cite{PRD21p772},
shift the central masses of the $S$ and $D$ charmonium
resonances to the positions indicated in the figure.
One may observe that all enhancements, with the exception
of the one above the $\Lambda_{c}^{+}\Lambda_{c}^{-}$ threshold,
correspond to the predicted central mass positions.
In the inset we show independent Breit-Wigner fits
to each of those resonances that had not yet been firmly determined
in previous work.}
\label{4Sto8S}
\end{figure}

While analysing the situation of the $X(4260)$ enhancement
in $\pi^{+}\pi^{-}J/\psi$, we furthermore found indications
in the data of Ref.~\cite{PRL95p142001}
for the existence
of several new $c\bar{c}$ resonances,
namely the $\psi (3D)$, $\psi (5S)$, $\psi (4D)$, $\psi (6S)$,
and $\psi (5D)$ \cite{ARXIV09044351}.
These resonances had been previously identified by us
\cite{EPL85p61002}
in data from the Belle Collaboration \cite{PRL101p172001},
which revealed the $X(4630)$ enhancement
in the reaction $e^{+}e^{-}\to\Lambda_{c}^{+}\Lambda_{c}^{-}$.
Here, in the data of Ref.~\cite{PRD79p092001},
we observe (see Fig.~\ref{LcLcb}) that
the enhancement at the $\Lambda_{c}^{+}\Lambda_{c}^{-}$ threshold
is much more modest, as compared to
the $\psi (5S)$ and $\psi (4D)$ signals,
than in Ref.~\cite{PRL101p172001} (see Fig.~\ref{LcLca}).
The reason is, in our philosophy, that the Belle Collaboration searched for
$\Lambda_{c}^{+}\Lambda_{c}^{-}$ pairs, which couple only modestly to
$c\bar{c}$ states because double quark-pair annihilation is required.
On the other hand, the $D^{\ast}\bar{D}^{\ast}$ pairs observed by BaBar only
need single $q\bar{q}$ creation.
Note that the first data point in Fig.~\ref{LcLcb} is not considered in
our curve describing the  $\Lambda_{c}^{+}\Lambda_{c}^{-}$ threshold
enhancement, since it appears to be due to the $\psi(3D)$ resonance (see
Fig.~\ref{4Sto8S}).

Also notice that the shape of the enhancement just above the
$\Lambda_{c}^{+}\Lambda_{c}^{-}$ threshold (see Fig.~\ref{LcLcb})
is very similar to that above the $D^{\ast}_{s}\bar{D}^{\ast}_{s}$
threshold (see Fig.~\ref{B4260b}). Moreover, each enhancement
carries two, more pronounced, $\psi$ resonances on its shoulder.

Upon inspecting the present BaBar \cite{PRD79p092001} data for the reaction
$e^{+}e^{-}\to D^{\ast}\bar{D}^{\ast}$ from the $\psi (4S)$ resonance
upwards, we find clear indications for eight more vector charmonium excitations
(see Fig.~\ref{4Sto8S}).

Although statistics is poor, albeit
the best at our disposal after several decades
of vary scarce data on charmonium spectroscopy,
we observe that all enhancements in the data of Ref.~\cite{PRD79p092001},
with the exception of the enhancement right above the
$\Lambda_{c}^{+}\Lambda_{c}^{-}$ threshold,
are in accurate agreement with the $c\bar{c}$ resonances
predicted by the model formulated in Ref.~\cite{PRD21p772},
using the parameters of Ref.~\cite{PRD27p1527}.

The signal for the $\psi (5D)$
is very poor in the present data \cite{PRD79p092001}.
However, it has been observed \cite{EPL85p61002}
in Belle data \cite{PRL101p172001}, namely at $\approx 5.29$ GeV,
and, furthermore, as a rather clear enhancement \cite{ARXIV09044351}
in BaBar data \cite{ARXIV08081543}, viz.\ at $\approx 5.30$ GeV.

One may wonder why the BaBar Collaboration has not stressed
the results presented here in Fig.~\ref{4Sto8S}.
Is it that statistics alone does not allow for any firm conclusions?
To a certain extent, we may even agree with such a point of view.
However, whereas each individual new resonance identified by us
has very poor statistics, the regular pattern of enhancements in
Fig.~\ref{4Sto8S} can hardly be just ``noise''. It is certainly true
that alternative, exotic models may very well be able to reproduce
the masses of some of these enhancements. But such bound-state approaches
are doomed to predict many other and lighter states as well, not observed
so far, apart from their manifest incapacity to describe scattering and
production data.

The results in Fig.~\ref{4Sto8S}, if confirmed, are of paramount importance
for hadronic physics, as they point in a direction very different from
what nowadays is considered common wisdom in meson spectroscopy, namely the
adequacy of a confining potential that rises linearly for increasing
distances. There can be absolutely no doubt about the dramatic failure of
such a potential in reproducing an approximately equidistant spectrum as
suggested by Fig.~\ref{4Sto8S}. In contrast, the resonance-spectrum
expansion (RSE) \cite{IJTPGTNO11p179} based on the harmonic oscillator
(HO), to be denoted by HORSE henceforth, turns out to be a very successful
approach to mesonic resonance spectra, by combining HO confinement
\cite{PRD21p772} with the nonperturbative effect of meson loops.

The data of Fig.~\ref{4Sto8S} appear to contradict, in particular,
the spin-orbit splittings as predicted in Ref.~\cite{PRD72p054026}.
In the latter model, the $S$-$D$ splittings for vector
$c\bar{c}$ states become smaller for higher radial excitations,
being only about 20 MeV for the $6D$-$7S$ splitting. From
Fig.~\ref{4Sto8S} we estimate this splitting to be
roughly five to ten times larger.
Now, in the HORSE, $S$-$D$ splittings are exactly zero
at the quenched level, but get generated by meson loops.
For the corresponding couplings,
the three-meson vertices determined in Ref.~\cite{ZPC21p291} are employed,
which involve the orbital and spin quantum numbers,
not only of the $c\bar{c}$ pair, but also of the mesons in the loops.
The resulting $S$-$D$ splittings come out very different then,
apart from the fact that the physical vector charmonium resonances naturally
appear as mixtures of $S$ and $D$ states.
We find that the combination dominated by the $D$ wave
at most shifts a few tens of MeVs from the corresponding bare level
(indicated by {\color{blue} X} in Fig.~\ref{4Sto8S}).
The dominantly $S$-wave combination shifts substantially more,
viz.\ some 100--200 MeV, depending on the precise locations of nearby
thresholds.
This pattern is, to some extent, systematically repeated
for higher radial excitations, which the present data seem to confirm.

In the following, we shall briefly discuss the experimental status
in the light, heavy-light, and heavy sectors, as well as some of the HORSE
achievements there.

In the light-quark sector, the experimental situation
is confusing, especially concerning the vector mesons.
The $\rho$(1250--1300), observed in many experiments, has no separate entry
in the PDG tables \cite{PLB667p1}, though some observations are included under
the $\rho(1450)$ \cite{PLB667p1}. Strikingly, a $\rho$(1250--1300) is
completely incompatible with models employing linear comfinement, e.g.\ the
semirelativistic approach of Godfrey and Isgur \cite{PRD32p189}.
However, this is not a reasonable justification for purging it from the PDG
tables.
A $\rho(1290)$ was predicted by an early version \cite{PRD27p1527}
of the HORSE.
Moreover, resonances with central masses
ranging from as low as 1430 MeV to as high as 1850 MeV are collected under
one entry, viz.\ the $\rho (1700)$ \cite{PLB667p1}.
A very similar and chaotic situation exists for the vector $\phi$ resonances
\cite{PLB667p1,PRD80p094011}.
Last but not least, the firmly established $K^\ast(1410)$ \cite{PLB667p1}
is also totally at odds with linear confinement, being almost
200~MeV lighter than predicted in Ref.~\cite{PRD32p189}.

The status of the light and intermediate scalar mesons is even more
controversial. Whereas the HORSE predicts five scalar nonets for masses
up to about 2.2 GeV \cite{AIPCP814p143}, it is stated in many publications
that there are more resonances observed than predicted
by theory, thereby sometimes referring to QCD,
or even, without further specification,
to the theory of strong interactions.
The connection between this theory and the HORSE is not at all clear so far.

In Ref.~\cite{PRD21p772}, the first version of the HORSE was proposed, as
a nonrelativistic Schr\"{o}dinger model for the amplitude
in non-exotic multi-channel meson-meson scattering,
which allows an exact solution in the form
of an analytic expression for the $S$ matrix.
Bound states and resonances are obtained
through the coupling of the two-meson system
to an HO, the oscillator frequency being
independent of flavor.
By fine-tuning the intensity of the coupling,
one can transform the oscillator spectrum into
the spectrum of mesons, for all possible flavor combinations
\cite{PRD27p1527}.

The very same model was then applied to the light scalar mesons
\cite{ZPC30p615}, with unchanged parameters. Thus, a
low-lying scalar nonet comprising the resonances $f_{0}(600)$
(alias $\sigma$), $K^{\ast}_{0}(800)$ (alias $\kappa$), $f_{0}(980)$,
and $a_{0}(980)$ was predicted, as dynamically generated poles owing
their very existence to the strong coupling of bare HO states to
$S$-wave meson-meson channels \cite{ZPC30p615}. Simultaneously, another
nonet is generated, consisting of the
$f_{0}(1370)$, $K^{\ast}_{0}(1430)$, $f_{0}(1500)$, and $a_{0}(1450)$.
These resonances stem directly from the bare $q\bar{q}$ states of the
HO spectrum. In total, for energies up to about 2.2~GeV, the model
predicts three nonets of light scalars that can be linked to bare states,
besides two dynamically generated nonets \cite{AIPCP1030p219}.

In the charm-strange sector, a single-channel version of the HORSE
successfully described the narrow scalar meson
$D_{s0}^{\ast}(2317)$ \cite{PLB667p1} below the $KD$ threshold, alternatively
as a dynamically generated  resonance \cite{PRL91p012003} or a strongly
shifted and distorted $c\bar{s}$ state \cite{MPLA19p1949}. In either
description, the $D_{s0}^{\ast}(2317)$ has $c\bar{s}$ and $DK$ components
of comparable magnitude. In a multichannel extension of the model
\cite{PRL97p202001}, the first radial excitation of the
$D_{s0}^{\ast}(2317)$ was predicted at about 2.85~GeV, with a width of some
50 MeV, being a good candidate for the $D_{sJ}(2860)$ \cite{PRL97p222001}.

In Ref.~\cite{PRD80p074001}, we showed that the $\Upsilon(10580)$ signal is a
consequence of the opening of the $B\bar{B}$ open-bottom channel,
rather than being due to a resonance pole of the $b\bar{b}$ propagator.
The true $\Upsilon (4S)$ is probably the state at 10.684~GeV observed by the
CLEO Collaboration \cite{PRL54p381}, back in 1985, which mysteriously never
made it to the PDG tables. In very recent BaBar data \cite{PRL102p012001},
this vector $b\bar{b}$ resonance can be observed again, now fitted
\cite{ARXIV09100967} with a Breit-Wigner mass of 10.735 GeV and a width of
38 MeV.

The level spacing of the bare quark-antiquark spectrum
in the HORSE is given by $\omega =0.19$ GeV,
independent of the flavors involved.
This feature stems from the Anti-De Sitter (AdS)
confinement solution for QCD,
which follows from Weyl conformal invariance \cite{NCA80p401}.
The latter solution has a further interesting property, namely that
the lowest-order potential-like term of the interaction
has the same form as the funnel-type potential deduced from lattice QCD.
However, notwithstanding this lowest-order term, the relativistic AdS spectrum
is exactly the same as that for the nonrelativistic HO
\cite{LNP211p331}.

The level spacing $2\omega =0.38$ GeV
can nicely be observed in Fig.~\ref{4Sto8S}.
If we take the mass of the charm quark from Ref.~\cite{PRD27p1527},
viz.\ 1.562 GeV, then the degenerate HO level
of the $\psi (4S)$ and $\psi (3D)$ comes out at
$2m_{c}+\omega (6+3/2)=4.549$ GeV,
with the next higher radial excitations at
4.929, 5.309, and 5.689 GeV.
These values are indicated by ({\color{blue} X})
on the horizontal axis of Fig.~\ref{4Sto8S}.
Meson loops then bring the various resonances
to their central masses and give them a hadronic decay width,
as foreseen back in 1980 \cite{PRD21p772}.
After three decades, this prediction is finally confirmed here.
Of course, in the meantime the HORSE has developed
into a more general formalism,
but the basic features have not changed much.
In particular, the quark masses and the oscillator frequency
have been kept at the values proposed in Ref.~\cite{PRD27p1527}.

\begin{table}[htbp]
\begin{center}
\begin{tabular}{||c||c|c||}
\hline\hline & & \\ [-7pt]
resonance & mass (GeV) & width (MeV)\\
& & \\ [-7pt]
\hline & & \\ [-7pt]
$\psi (4S)$ & 4.42 \cite{PLB667p1} & 62 \cite{PLB667p1}\\
$\psi (3D)$ & $\approx$4.55 \cite{ARXIV09044351} &
$\approx$50 \cite{ARXIV09044351}\\
$\psi (5S)$ & 4.78 \cite{PRD80p074001} &
55 \cite{PRD80p074001}\\
$\psi (4D)$ & 4.87 \cite{PRD80p074001} &
60 \cite{PRD80p074001}\\
$\psi (6S)$ & 5.09 & 55\\
$\psi (5D)$ & $\approx$5.30\cite{ARXIV09044351} &
$\approx$70\cite{ARXIV09044351}\\
$\psi (7S)$ & 5.44 & 44\\
$\psi (6D)$ & 5.66 & 53\\
$\psi (8S)$ & 5.91 & 93\\ [5pt]
\hline\hline
\end{tabular}
\end{center}
\caption[]{\small
Breit-Wigner masses and widths
for the charmonium vector resonances
indicated in Fig.~\ref{4Sto8S}.
The resonance parameters for the $\psi (4S)$ are taken from
Ref.~\cite{PLB667p1},
while those for the $\psi (3D)$, and $\psi (5D)$, are deduced from
Ref.~\cite{ARXIV09044351}, where statistics was slightly better.
For the $\psi (5S)$, and $\psi (4D)$,
we take the resonance parameters from Ref.~\cite{PRD80p074001}.
The remaining resonances are indepently fitted,
as shown in the inset of Fig.~\ref{4Sto8S}.}
\label{masswidth}
\end{table}
In Table~\ref{masswidth}, we give the resonance parameters
for the $\psi (4S)$ and the eight charmonium vector
resonances observed in Fig.~\ref{4Sto8S}, with the proviso that these
numbers may very well turn out to be significantly corrected by future
data, due to the present low statistics.

In conclusion,
the recent data for the reaction $e^{+}e^{-}\to D^{\ast}\bar{D}^{\ast}$
published BaBar \cite{PRD79p092001}
show that the $X(4260)$ and the $Y(4660)$
are not to be associated with resonance poles
of the $c\bar{c}$ propagator.
Moreover, the same data provide evidence for the new
charmonium vector states
$\psi (3D)$, $\psi (5S)$, $\psi (4D)$, $\psi (6S)$,
$\psi (5D)$, $\psi (7S)$, $\psi (6D)$, and $\psi (8S)$.
These findings supports the HO model for
the bare quark-antiquark propagator \cite{NCA80p401},
and for the way meson loops are accounted for,
in a nonperturbative fashion, so as
to obtain a unitary scattering matrix \cite{PRD21p772}
as well as the corresponding production amplitudes \cite{AP323p1215}.

Finally, we should emphasize that no detailed coupled-channel calculation
has been carried out in the present analysis, as would be possible in principle
within the HORSE framework, along the lines worked out by us in many
other papers and applied to a variety of mesonic resonances. Such a
calculation would be a huge endeavor though, in view of the proliferation
of decay channels for the highly excited $c\bar{c}$ states described here,
many of which involving resonances themselves. Nevertheless, the striking
regularities manifest in the present charmonium data, which are perfectly
compatible with excitation levels and coupled-channel mass shifts
successfully determined in the HORSE for many other mesons, make us
confident in the reliability of our analysis.
\vspace{0.3cm}

{\bf Acknowledgments}:
We are grateful for the precise measurements
and data analyses of the BaBar Collaboration, which
made the present analysis possible.
We also thank Jorge Segovia for carrying out the Breit-Wigner fits
shown in the inset of Fig.~\ref{4Sto8S}.
This work was supported in part by the {\it Funda\c{c}\~{a}o para a
Ci\^{e}ncia e a Tecnologia} \/of the {\it Minist\'{e}rio da Ci\^{e}ncia,
Tecnologia e Ensino Superior} \/of Portugal, under contract
CERN/\-FP/\-109307/\-2009.

\newcommand{\pubprt}[4]{#1 {\bf #2}, #3 (#4)}
\newcommand{\ertbid}[4]{[Erratum-ibid.~#1 {\bf #2}, #3 (#4)]}
\def\AIPCP{AIP Conf.\ Proc.}
\def\AP{Ann.\ Phys.}
\def\EPL{Europhys.\ Lett.}
\def\IJTPGTNO{Int.\ J.\ Theor.\ Phys.\ Group Theor.\ Nonlin.\ Opt.}
\def\LNP{Lect.\ Notes Phys.}
\def\MPLA{Mod.\ Phys.\ Lett.\ A}
\def\NCA{Nuovo Cim.\ A}
\def\PLB{Phys.\ Lett.\ B}
\def\PRD{Phys.\ Rev.\ D}
\def\PRL{Phys.\ Rev.\ Lett.}
\def\ZPC{Z.\ Phys.\ C}


\begin{thebibliography}{39}
\bibitem{PRD79p092001}
B.~Aubert  [The BABAR Collaboration],
\pubprt{\PRD}{79}{092001}{2009}
[arXiv:0903.1597 [hep-ex]].
%%CITATION = PHRVA,D79,092001;%%

\bibitem{PLB667p1}
C.~Amsler {\it et al.} \/[Particle Data Group Collaboration],
\pubprt{\PLB}{667}{1}{2008}.
%%CITATION = PHLTA,B667,1;%%

\bibitem{PRL95p142001}
B.~Aubert {\it et al.}  [BABAR Collaboration],
\pubprt{\PRL}{95}{142001}{2005}
[arXiv:hep-ex/0506081].
%%CITATION = HEP-EX 0506081;%%

\bibitem{PRL101p172001}
G.~Pakhlova {\it et al.}  [Belle Collaboration],
\pubprt{\PRL}{101}{172001}{2008}
[arXiv:0807.4458 [hep-ex]].
%%CITATION = PRLTA,101,172001;%%

\bibitem{PRL96p162003}
T.~E.~Coan {\it et al.}  [CLEO Collaboration],
\pubprt{\PRL}{96}{162003}{2006}
[arXiv:hep-ex/0602034].
%%CITATION = HEP-EX 0602034;%%

\bibitem{PRL99p182004}
C.~Z.~Yuan {\it et al.}  [Belle Collaboration],
\pubprt{\PRL}{99}{182004}{2007}
[arXiv:0707.2541 [hep-ex]].
%%CITATION = PRLTA,99,182004;%%

\bibitem{PRL98p212001}
B.~Aubert {\it et al.}  [BABAR Collaboration],
\pubprt{\PRL}{98}{212001}{2007}
[arXiv:hep-ex/0610057].
%%CITATION = PRLTA,98,212001;%%

\bibitem{PRL99p142002}
X.~L.~Wang {\it et al.}  [Belle Collaboration],
\pubprt{\PRL}{99}{142002}{2007}
[arXiv:0707.3699 [hep-ex]].
%%CITATION = PRLTA,99,142002;%%

\bibitem{AP323p1215}
E.~van Beveren and G.~Rupp,
\pubprt{\AP}{323}{1215}{2008}
[arXiv:0706.4119].
%%CITATION = ARXIV:0706.4119;%%

\bibitem{PRD21p772}
E.~van Beveren, C.~Dullemond, and G.~Rupp,
\pubprt{\PRD}{21}{772}{1980}
\ertbid{\ D}{22}{787}{1980}.
%%CITATION = PHRVA,D21,772;%%

\bibitem{PRD32p189}
S.~Godfrey and N.~Isgur,
\pubprt{\PRD}{32}{189}{1985}.
%%CITATION = PHRVA,D32,189;%%

\bibitem{IJTPGTNO11p179}
E.~van Beveren and G.~Rupp,
\pubprt{\IJTPGTNO}{11}{179}{2006}
[arXiv:hep-ph/0304105].
%%CITATION = HEP-PH 0304105;%%

\bibitem{EPL81p61002}
E.~van Beveren and G.~Rupp,
\pubprt{\EPL}{81}{61002}{2008}
[arXiv:0710.5823 [hep-ph]].
%%CITATION = ARXIV:0710.5823;%%

\bibitem{EPL84p51001}
M.~R.~Pennington and D.~J.~Wilson,
\pubprt{\EPL}{84}{51001}{2008}.
%%CITATION = EULEE,84,51001;%%

\bibitem{PRD35p1633}
K.~L.~Au, D.~Morgan and M.~R.~Pennington,
\pubprt{\PRD}{35}{1633}{1987}.
%%CITATION = PHRVA,D35,1633;%%

\bibitem{EPL84p51002}
E.~van Beveren and G.~Rupp,
\pubprt{\EPL}{84}{51002}{2008}.
%%CITATION = EULEE,84,51002;%%

\bibitem{HEPPH0605317}
E.~van Beveren and G.~Rupp,
arXiv:hep-ph/0605317.
%%CITATION = HEP-PH 0605317;%%

\bibitem{ARXIV08111755}
E.~van Beveren and G.~Rupp,
in Proceedings {\it Bled Workshops in Physics},
Vol.~9, no.~1, pp 26-29 (2008)
[arXiv:0811.1755 [hep-ph]].
%%CITATION = ARXIV:0811.1755;%%

\bibitem{ARXIV09044351}
E.~van Beveren and G.~Rupp,
arXiv:0904.4351 [hep-ph].
%%CITATION = ARXIV:0904.4351;%%

\bibitem{PRD79p111501R}
E.~van Beveren and G.~Rupp,
\pubprt{\PRD}{79}{111501R}{2009}
[arXiv:0905.1595 [hep-ph]].
%%CITATION = ARXIV:0905.1595;%%

\bibitem{EPL85p61002}
E.~van~Beveren, X.~Liu, R.~Coimbra, and G.~Rupp,
\pubprt{\EPL}{85}{61002}{2009}
[arXiv:0809.1151 [hep-ph]].
%%CITATION = EULEE,85,61002;%%

\bibitem{PRD27p1527}
E.~van Beveren, G.~Rupp, T.~A.~Rij\-ken, and C.~Dullemond,
\pubprt{\PRD}{27}{1527}{1983}.
%%CITATION = PHRVA,D27,1527;%%

\bibitem{ARXIV08081543}
B.~Aubert  [BaBar Collaboration],
arXiv:0808.1543 [hep-ex].
%%CITATION = ARXIV:0808.1543;%%

\bibitem{PRD72p054026}
T.~Barnes, S.~Godfrey and E.~S.~Swanson,
\pubprt{\PRD}{72}{054026}{2005}
[arXiv:hep-ph/0505002].
%%CITATION = PHRVA,D72,054026;%%

\bibitem{ZPC21p291}
E.~van Beveren,
\pubprt{\ZPC}{21}{291}{1984}
[arXiv:hep-ph/0602246].
%%CITATION = ZEPYA,C21,291;%%

\bibitem{PRD80p094011}
S.~Coito, G.~Rupp and E.~van Beveren,
\pubprt{\PRD}{80}{094011}{2009}
[arXiv:0909.0051 [hep-ph]].
%%CITATION = PHRVA,D80,094011;%%

\bibitem{AIPCP814p143}
E.~van Beveren, F.~Kleefeld and G.~Rupp,
\pubprt{\AIPCP}{814}{143}{2006}
[arXiv:hep-ph/0510120].
%%CITATION = HEP-PH 0510120;%%

\bibitem{ZPC30p615}
E.~van Beveren, T.~A.~Rij\-ken, K.~Metzger, C.~Dullemond, G.~Rupp and
J.~E.~Ribeiro,
\pubprt{\ZPC}{30}{615}{1986}
[arXiv:0710.4067 [hep-ph]].
%%CITATION = ZEPYA,C30,615;%%

\bibitem{AIPCP1030p219}
E.~van Beveren and G.~Rupp,
\pubprt{\AIPCP}{1030}{219}{2008}
[arXiv:0804.2573 [hep-ph]].
%%CITATION = ARXIV:0804.2573;%%

\bibitem{PRL91p012003}
E.~van Beveren and G.~Rupp,
\pubprt{\PRL}{91}{012003}{2003}
[arXiv:hep-ph/0305035].
%%CITATION = HEP-PH 0305035;%%

\bibitem{MPLA19p1949}
E.~van Beveren and G.~Rupp,
\pubprt{\MPLA}{19}{1949}{2004}
[arXiv:hep-ph/0406242].
%%CITATION = HEP-PH 0406242;%%

\bibitem{PRL97p202001}
E.~van Beveren and G.~Rupp,
\pubprt{\PRL}{97}{202001}{2006}
[arXiv:hep-ph/0606110].
%%CITATION = HEP-PH 0606110;%%

\bibitem{PRL97p222001}
B.~Aubert  [BABAR Collaboration],
\pubprt{\PRL}{97}{222001}{2006}
[arXiv:hep-ex/0607082].
%%CITATION = HEP-EX 0607082;%%

\bibitem{PRD80p074001}
E.~van Beveren and G.~Rupp,
\pubprt{\PRD}{80}{074001}{2009}
[arXiv:0908.0242 [hep-ph]].
%%CITATION = ARXIV:0908.0242;%%

\bibitem{PRL54p381}
D.~Besson {\it et al.} [CLEO Collaboration],
\pubprt{\PRL}{54}{381}{1985}.
%%CITATION = PRLTA,54,381;%%

\bibitem{PRL102p012001}
B.~Aubert  [BaBar Collaboration],
\pubprt{\PRL}{102}{012001}{2009}
[arXiv:0809.4120 [hep-ex]].
%%CITATION = PRLTA,102,012001;%%

\bibitem{ARXIV09100967}
E.~van Beveren and G.~Rupp,
arXiv:0910.0967 [hep-ph].
%%CITATION = ARXIV:0910.0967;%%

\bibitem{NCA80p401}
C.~Dullemond, T.~A.~Rij\-ken and E.~van Beveren,
\pubprt{\NCA}{80}{401}{1984}.
%%CITATION = NUCIA,A80,401;%%

\bibitem{LNP211p331}
E.~van Beveren, T.~A.~Rij\-ken, C.~Dullemond and G.~Rupp,
\pubprt{\LNP}{211}{331}{1984}.
%%CITATION = LNPHA,211,331;%%
\end{thebibliography}
\end{document}